\documentclass[aps,prd,superscriptaddress,nofootinbib,floats,preprint,floats,floatfix]{revtex4-1}
\usepackage{bm}
\usepackage{indentfirst}
\usepackage{amsmath}
\usepackage{graphicx}
\usepackage{float}
\usepackage{amssymb}
\usepackage{subfigure}
\usepackage{psfrag}
\usepackage{hyperref}
\usepackage{tensor}
\usepackage{braket}
\usepackage{multirow}
\usepackage{verbatim}
\usepackage{ulem}
\usepackage{cancel}

\usepackage[dvipsnames, svgnames, x11names]{xcolor}
\hypersetup{
	colorlinks=true,
	linkcolor=red,
	citecolor=blue,
} 

\allowdisplaybreaks[1]
\usepackage[utf8]{inputenc}

\begin{document}

\title{Scalar induced gravitational waves in symmetric teleparallel gravity with a parity-violating term}
\author{Fengge Zhang}
\email{zhangfg5@mail.sysu.edu.cn}
\affiliation{School of Physics and Astronomy, Sun Yat-sen University, Zhuhai 519082, China}
\author{Jia-Xi Feng}
\email{fengjx57@mail2.sysu.edu.cn}
\affiliation{School of Physics and Astronomy, Sun Yat-sen University, Zhuhai 519082, China}
\author{Xian Gao}
\email{gaoxian@mail.sysu.edu.cn (corresponding author)}
\affiliation{School of Physics and Astronomy, Sun Yat-sen University, Zhuhai 519082, China}

\begin{abstract}
 Gravitational waves (GWs) are useful to test gravitational theories and to probe the physics in the early universe. In this paper, we investigate the scalar induced gravitational waves (SIGWs) in symmetric teleparallel gravity with a parity-violating term. The presence of the parity-violating term leads to the velocity birefringence effect of the SIGWs. However, after taking into account the observational constraints on the speed of GWs, the contribution from the parity-violating term to SIGWs is negligible. Nevertheless, the contribution to SIGWs from the perturbations of the connection can be significant, and results in a multipeak structure in the energy density of SIGWs. This feature makes the symmetric teleparallel gravity distinguishable from the general relativity.
\end{abstract}

\maketitle


\section{Introduction}
The detection of gravitational waves (GWs) by the Laser Interferometer Gravitational-Wave Observatory (LIGO) scientific collaboration and Virgo collaboration \cite{Abbott:2016nmj,Abbott:2016blz,Abbott:2017gyy,TheLIGOScientific:2017qsa,Abbott:2017oio,Abbott:2017vtc,LIGOScientific:2018mvr,Abbott:2020khf,Abbott:2020uma,LIGOScientific:2020stg} opens a new window to probe the nature of gravity in the strong gravitational field and nonlinear regime. Although the observation from cosmic microwave background (CMB) constrains the power spectrum of primordial curvature perturbation to be $\mathcal{A}_{\zeta} \sim \mathcal{O}(10^{-9})$ on  large scales \cite{Akrami:2018odb}, it can be as large as $\mathcal{A}_{\zeta} \sim \mathcal{O}(10^{-2})$ on small scales \cite{Sato-Polito:2019hws,Lu:2019sti}. Such large scalar perturbation will induce the generation of gravitational waves, which are dubbed of the scalar induced gravitational waves (SIGWs), due to the nonlinear interactions between the scalar and tensor perturbations \cite{Ananda:2006af,Saito:2008jc,Orlofsky:2016vbd,Nakama:2016gzw,Wang:2016ana,Kohri:2018awv,Espinosa:2018eve,Kuroyanagi:2018csn,Domenech:2019quo,Fumagalli:2020nvq,Lin:2020goi,Domenech:2020kqm,Lu:2020diy,Domenech:2021ztg,Zhang:2021vak,Wang:2021djr,Zhang:2021rqs,Zhang:2021rqs,Yi:2022ymw}. The SIGWs can be large enough to be detected by the space-based GW observatories, such as  Laser Interferometer Space Antenna (LISA) \cite{Danzmann:1997hm,LISA:2017pwj}, TianQin \cite{Luo:2015ght} and Taiji \cite{Hu:2017mde}, as well as by the Pulsar Timing Array (PTA) \cite{Kramer:2013kea,Hobbs:2009yy,McLaughlin:2013ira,Hobbs:2013aka} and the Square Kilometer Array (SKA) \cite{Moore:2014lga} in the future. 

Discrete symmetries, such as parity, play an important role in modern physics. 
While the parity is known to be violated in weak interactions \cite{Lee:1956qn,Wu:1957my}, one may wonder whether this symmetry violation exists in gravitational interactions and/or in the early universe as well. The parity-violating (PV) gravitational theories are generally predicted in quantum gravity theories such as the superstring theory and M-theory \cite{Green:1984sg,Witten:1984dg}. The recent hints of parity-violation in our universe from galaxy trispectrum and the CMB E/B cross-correlation also have attracted much attention \cite{Philcox:2022hkh,Hou:2022wfj,Minami:2020odp,Eskilt:2022cff}. The parity-violating scalar trispectrum was also studied in \cite{Liu:2019fag,Niu:2022fki,Cabass:2022rhr,Creque-Sarbinowski:2023wmb} recently.

The simplest PV term in the Riemannian geometry is the Chern-Simons (CS) term, which is quadratic in the Riemann tensor. The CS gravity was first proposed in \cite{Jackiw:2003pm} in four-dimensional spacetime, and later  extensively  studied in cosmology, GWs, and primordial non-Gaussianity \cite{Lue:1998mq,Satoh:2007gn,Saito:2007kt,Satoh:2007gn,Alexander:2009tp,Yunes:2010yf,Gluscevic:2010vv,Yunes:2010yf,Myung:2014jha,Kawai:2017kqt,Nair:2019iur,Nishizawa:2018srh,Odintsov:2022hxu,Bartolo:2017szm,Bartolo:2018elp}. Besides CS gravity, the PV gravity models with Lorentz breaking, such as Ho\v{r}ava gravity \cite{Horava:2009uw}, the PV higher derivative gravity \cite{Crisostomi:2017ugk} and the PV spatially covariant gravity \cite{Gao:2019liu,Hu:2021bbo,Hu:2021yaq} have also been proposed. In these Lorentz breaking PV gravity models, the chiral GWs have been  studied extensively \cite{Takahashi:2009wc,Myung:2009ug,Wang:2012fi,Zhu:2013fja,Cannone:2015rra,Zhao:2019szi,Zhao:2019xmm,Qiao:2019hkz,Qiao:2019wsh,Qiao:2021fwi,Gong:2021jgg}, wherein  interesting features of GWs were revealed, notably including phenomena such as the velocity and amplitude birefringence.

Recently, there are also interests in gravity theories based on non-Riemannian geometry. In particular, teleparallel gravity, which is with non-metricity tensor $Q_{\rho\mu\nu} = \nabla_{\rho} g_{\mu\nu}$ and/or torsion, were proposed and attracted much attention \cite{Nieh:1981ww,Chatzistavrakidis:2020wum,Cai:2021uup,Wu:2021ndf,Langvik:2020nrs,Li:2020xjt,Li:2021wij,Rao:2021azn,Li:2021mdp,Li:2022mti,Li:2022vtn,Hohmann:2020dgy,Bombacigno:2021bpk,Iosifidis:2020dck,Hohmann:2022wrk,Conroy:2019ibo,Iosifidis:2021bad,Pagani:2015ema,Chen:2022wtz,Li:2022vtn,Li:2022mti,Gialamas:2022xtt,Gialamas:2023emn,Battista:2021rlh,Battista:2022hmv,Papanikolaou:2022hkg,Tzerefos:2023mpe}. 
Similar to the CS gravity, the simplest PV term built of the non-metricity tensor is $\widetilde{Q}Q \equiv  \varepsilon_{\mu\nu\rho\sigma}Q^{\mu\nu}_{\phantom{\mu\nu}\lambda}Q^{\rho\sigma\lambda}$, which is quadratic in the non-metricity tensor. The simplest symmetric teleparallel gravity with PV was constructed by appending this term to the symmetric teleparallel equivalent Einstein-Hilbert action, of which the linear cosmological perturbation has also been studied \cite{Conroy:2019ibo,Li:2021mdp}. 
It was shown that the PV term has no contribution to the background evolution or the linear scalar perturbations.

The SIGWs in CS gravity have been studied in \cite{Zhang:2022xmm,Feng:2023veu} recently.
The purpose of this work is to perform a similar study of the SIGWs in the symmetric teleparallel gravity with PV terms.
However, when the nonlinear perturbations are taken into account, the above simplest PV symmetric teleparallel gravity may be inconsistent. 
Intuitively, the theory contains extra scalar degrees of freedom due to the PV term, which however do not show themselves on the linear order around a homogeneous and isotropic background.
This is reminiscent of the so-called strong coupling problem in the study of the Ho\v{r}ava gravity \cite{Blas:2009yd,Charmousis:2009tc,Blas:2009qj,Papazoglou:2009fj,Blas:2009ck} and $f(T)$ gravity \cite{Ferraro:2018tpu,Li:2011rn,Izumi:2012qj,Hu:2023juh,Golovnev:2020zpv}.

As we will demonstrate in this paper, the simplest PV symmetric teleparallel gravity model suffers from such a strong coupling problem. Specifically, the scalar perturbations from the connection do not have the linear equations of motion of their own, which arise in the equation of motion of the SIGWs. To avoid this  problem, we modify the symmetric teleparallel equivalent Einstein-Hilbert action by considering a general linear combination of quadratic monomials of the non-metricity tensor. 
We then obtain the equations of motion of perturbations from connection, as well as find the solution of the perturbations from connection during the radiation-dominated era. Based on these results, we will calculate the contribution from the PV term as well as from the scalar perturbations of connection to the energy density of SIGWs in our model, respectively.

This paper is organized as follows. In section \ref{sec2}, we introduce the symmetric teleparallel gravity with a simple PV term. In section \ref{sec3}, we give the equations of motion for both the background evolution and the linear scalar perturbations, which we then solve during the radiation-dominated era. In section \ref{sec4}, we derive the equation of motion of SIGWs. In section \ref{sec5}, we calculate the power spectra of the SIGWs. In order to analyze the feature of SIGWs, we compute the energy density of SIGWs with the monochromatic power spectrum of primordial curvature perturbation. Our results are summarized in section \ref{seccon}. The quadratic action of linear scalar perturbations and the analytic part of the integral kernel are included in appendices \ref{AC2} and \ref{kernel}, respectively.

\section{The symmetric teleparallel gravity with a parity-violating term}\label{sec2}

In symmetric teleparallel gravity, the affine connection is assumed to be free of the curvature and torsion, i.e.,
\begin{equation}
R^{\mu}_{\ \nu\rho\sigma}=\partial_{\rho}\Gamma^{\mu}_{\ \nu\sigma}-\partial_{\sigma}\Gamma^{\mu}_{\ \nu\rho}+\Gamma^{\mu}_{\ \alpha\rho}\Gamma^{\alpha}_{\ \nu\sigma}-\Gamma^{\mu}_{\ \alpha\sigma}\Gamma^{\alpha}_{\ \nu\rho}=0,
\end{equation}
\begin{equation}
T^{\mu}_{\ \nu\rho}=\Gamma^{\mu}_{\ \rho\nu}-\Gamma^{\mu}_{\ \nu\rho}=0.
\end{equation}
The gravitational effects are encoded in the non-metricity tensor, which is defined as
\begin{equation}
Q_{\rho\mu\nu}=\nabla_{\rho}g_{\mu\nu}=\partial_{\rho}g_{\mu\nu}-\Gamma^{\sigma}_{\ \rho\mu}g_{\sigma\nu}-\Gamma^{\sigma}_{\ \rho\nu}g_{\sigma\mu},
\end{equation}
where $g_{\mu\nu}$ is the spacetime metric, $\nabla$ represents the covariant derivative. 

With the condition of vanishing curvature and torsion tensors, the coefficients of the connection take the following general form \cite{BeltranJimenez:2017tkd,DAmbrosio:2020nqu}
\begin{equation}\label{cy}
\Gamma^{\rho}_{\ \mu\nu}=\frac{\partial x^{\rho}}{\partial \xi^{\sigma}}\partial_{\mu}\partial_{\nu}\xi^{\sigma},
\end{equation}
where $\xi^{\mu}(x)$ are four general scalar fields.
If we choose $\xi^{\mu}(x)=x^{\mu}$, then $\Gamma^{\rho}_{\ \mu\nu}=0$. This is the so-called ``coincident gauge,'' which has been extensively used in the study of symmetric teleparallel gravity in order to simplify the calculation. In this paper, we do not  take the coincident gauge, because it may not be compatible with the commonly used conventional parametrization for metric when dealing with cosmological perturbations \cite{Zhao:2021zab,Li:2021mdp}.

Consider the following action
\begin{equation}\label{action}
S_g=\int \mathrm{d}^4x\sqrt{-g}\left(\frac{\mathbb{Q}}{2}-g(\varphi) \widetilde{Q}Q\right)+\int \mathrm{d}^4x\sqrt{-g}\left(\frac{1}{2}g^{\mu\nu}\partial_{\mu}\varphi\partial_{\nu}\varphi-V(\varphi)\right),
\end{equation}
where 
$\mathbb{Q}=P^{\alpha}_{\ \mu\nu}Q_{\alpha}^{\ \mu\nu} $ \cite{BeltranJimenez:2017tkd}, with
\begin{equation}
P^{\alpha}_{\ \mu\nu}=c_1Q^{\alpha}_{\ \mu\nu}+c_2Q_{(\mu\ \nu)}^{\ \ \alpha}+c_3Q^{\alpha}g_{\mu\nu}+c_4\delta^{\alpha}_{\ (\mu}\tilde{Q}_{\nu)}+\frac{c_5}{2}\left(\Tilde{Q}^{\alpha}g_{\mu\nu}+\delta^{\alpha}_{\ (\mu}Q_{\nu)}\right),
\end{equation}
where $c_{1},\cdots,c_{5}$ are constants, and
\begin{equation}
Q_{\mu}=Q_{\mu\ \alpha}^{\ \alpha},\ \ \tilde{Q}^{\mu}=Q_{\alpha}^{\ \mu\alpha}.
\end{equation}
The PV term is represented as \cite{Conroy:2019ibo}
\begin{equation}
\widetilde{Q}Q=\varepsilon^{\mu\nu\rho\sigma}Q_{\mu\nu\alpha}Q_{\rho\sigma}^{\ \ \alpha},
\end{equation}
where $\varepsilon^{\mu\nu\rho\sigma}=\epsilon^{\mu\nu\rho\sigma}/\sqrt{-g}$ is the Levi-Civita tensor, with $\epsilon^{\mu\nu\rho\sigma}$  the antisymmetric symbol.
In Eq. (\ref{action}) the scalar field effectively describes matter content in the universe.

By choosing the values of the parameters in the action \eqref{action} to be
\begin{equation}\label{ps}
c_1=-1/4,\ c_2=1/2,\ c_3=1/4,\ c_4=0,\ c_5=-1/2,
\end{equation}
the expression for $\mathbb{Q}$ becomes
\begin{equation}
	\mathbb{Q}=\frac{1}{4}Q_{\rho\mu\nu}Q^{\rho\mu\nu}-\frac{1}{2}Q_{\rho\mu\nu}Q^{\mu\nu\rho}-\frac{1}{4}Q^{\alpha}Q_{\alpha}+\frac{1}{2}\Bar{Q}^{\alpha}Q_{\alpha}=-\overcirc{R}-\overcirc{\nabla}_{\alpha}\left(Q^{\alpha}-\Bar{Q}^{\alpha}\right),
\end{equation}
which  corresponds to the teleparallel equivalent Einstein-Hilbert Lagrangian.
Here, $\overcirc{R}$ is constructed with the metric $g_{\mu\nu}$, and $\overcirc{\nabla}$ is metric-compatible covariant derivative.
In this case, the linear cosmological perturbations were studied in \cite{Li:2021mdp}. 
However, in the next section, we will show that this model suffers from the strong coupling problem at nonlinear orders, which can be avoided by choosing a suitable parameter set instead of \eqref{ps}.

\section{The cosmological perturbations}\label{sec3}
In this section, we study the evolution of background  and the linear scalar cosmological perturbations.

Consider the spatially flat Friedmann-Robertson-Walker (FRW) background with small perturbations around it, the metric under the Newtonian gauge is
\begin{equation}\label{metric}
\mathrm{d}s^2=a^2\left\{(1+2\phi+2\phi^2)\mathrm{d}\tau^2-\left[(1-2\psi+2\psi^2)\delta_{ij}+h_{ij}+\frac{1}{2}h_{ik}h^{k}_{\ j}\right]\mathrm{d}x^i\mathrm{d}x^j\right\},
\end{equation}
up to the second order in perturbations $\phi$, $\psi$ and $h_{ij}$,
and the components of the inverse metric are
\begin{gather}
g^{00}=\frac{1}{a^2}\left(1-2\phi+2\phi^2\right),\ \ g^{0i}=0, \\ g^{ij}=-\frac{1}{a^2}\left[\left(1+2\psi+2\psi^2\right)\delta^{ij}-h^{ij}-4\psi h^{ij}+\frac{1}{2}h^i_{\ l}h^{lj}+8\psi^2 h^{ij}\right].
\end{gather}
Note that for our purpose to evaluate the SIGWs, only quadratic action for the scalar and tensor perturbations as well as cubic action involving two scalar and one tensor perturbation modes are needed. 
Therefore, in the above expression for $g^{ij}$ we have kept only the cubic term $\psi^2h^{ij}$ for notational simplicity.
We also have
\begin{equation}
\sqrt{-g}=a^4\left(1+\phi-3\psi+\frac{1}{2}\phi^2-3\phi\psi+\frac{9}{2}\psi^2\right).
\end{equation}

From the Eq. \eqref{cy}, the components of the connection are fully determined by four scalar fields $\xi^{\mu}$, thus we can view the four scalar fields $\xi^{\mu}$ and metric $g_{\mu\nu}$ as fundamental variables in symmetric teleparallel gravity. At the background level, we can take the background variables that are related to the connection as $\bar{\xi}^{\mu}=x^{\mu}$, where $x^{\mu}$ are spacetime coordinates. Although this choice is dubbed the ``unitary gauge" in the literature \cite{BeltranJimenez:2022azb,Blixt:2023kyr}, it is actually the solution for the background configurations of the four scalar fields. While in the perturbed universe, we introduce small quantities $\delta \xi^{\mu}$ to represent the perturbation of the scalar fields $\xi^{\mu}$, thus we have the perturbed scalar fields $\xi^{\mu}=\bar{\xi}^{\mu}+\delta \xi^{\mu}=x^{\mu}+\delta\xi^{\mu}$. We further decompose $\delta\xi^{\mu}$ as $\delta\xi^{\mu}=\{C,\partial^{i}D\}$, where $C$ and $D$ are scalar perturbations. With these settings, the components of the connection can be expressed as
\begin{equation}
\Gamma^{\rho}_{\ \mu\nu}=\partial_{\mu}\partial_{\nu}\delta\xi^{\rho}-\partial_{\sigma}\delta\xi^{\rho}\partial_{\mu}\partial_{\nu}\delta\xi^{\sigma},
\end{equation}
up to the second order.

We split the scalar field $\varphi$ to be $\bar{\varphi} + \delta\varphi$, where $\bar{\varphi}(t)$ is the background value and $\delta\varphi$ is the perturbation.

\subsection{The EOMs of background}

By expanding the action \eqref{action} to the linear order in perturbations, we obtain the following action for the perturbations
\begin{equation}\label{ac1}
\begin{split}
S^{(1)}=&\int \mathrm{d}^3x \mathrm{d}\tau a^2\left[\left(2\mathcal{C}_1\mathcal{H}^2-\frac{1}{2}(\varphi')^2-a^2V\right)\phi-2\mathcal{C}_2\mathcal{H}\phi'+6\mathcal{C}_3\mathcal{H}\psi'\right.\\&\left.+3\left(2\mathcal{C}_1\mathcal{H}^2-\frac{1}{2}(\varphi')^2+a^2V\right)\psi-a^2V_{\varphi}\delta\varphi+\varphi'\delta\varphi'+2\mathcal{C}_2\mathcal{H}C''\right],
\end{split}
\end{equation}
where a prime denotes derivative with respect to the conformal time $\tau$, and 
\begin{gather}
\label{C1}
\mathcal{C}_1=4c_1+c_2+16c_3+c_4+4c_5,\\ 
\label{C2}
\mathcal{C}_2=2c_1+2c_2+8c_3+2c_4+5c_5, \\ 
\label{C3}
\mathcal{C}_3=2c_1+8c_3+c_5.
\end{gather}
Here and in what follows, we denote $\varphi$ the background value for the scalar field for simplicity.

Varying the above action \eqref{ac1} with respect to the perturbations $\phi$, $\psi$, $\delta\varphi$, and $C$, we obtain the equations of motion (EOMs) for the background 
\begin{gather}
\label{p11}2\left(\mathcal{C}_1+2\mathcal{C}_2\right)\mathcal{H}^2+2\mathcal{C}_2\mathcal{H}'=\frac{1}{2}(\varphi')^2+a^2V,\\
\label{p12}2\left(\mathcal{C}_1-2\mathcal{C}_3\right)\mathcal{H}^2-2\mathcal{C}_3\mathcal{H}'=\frac{1}{2}(\varphi')^2-a^2V,\\
\label{p13}\varphi''+2\mathcal{H}\varphi'+a^2V_{\varphi}=0,\\
\label{p14}\left(a^2\mathcal{H}\right)''=0 ~ \left(\mathcal{C}_2\neq 0\right).
\end{gather}
From the above EOMs (\ref{p11})-(\ref{p14}), we observe that the evolution of the background is unaffected by the PV term, as expected. 
As a consistency check, once we choose the parameter sets as \eqref{ps}, namely, $\mathcal{C}_1=3/2,\mathcal{C}_2=0$, and $\mathcal{C}_3=1$, the above equations of motion are the same as those in GR.
It is interesting to note that in the case of $\mathcal{C}_2\neq 0$, (\ref{p14}) acts as  an extra constraint equation for the background.

\subsection{The EOMs of the linear scalar perturbations}

In order to get the equations of motion for the linear perturbations, we expand the action \eqref{action} to the quadratic order in perturbations, which is tedious and can be found in Appendix \ref{AC2}. \footnote{Note that the form of the quadratic action \eqref{L2} will change by performing integrations by parts.}  

Varying the quadratic action \eqref{L2} with respect to the scalar perturbations $\phi,~\psi,~\delta\varphi,~C$ and $D$, we can obtain the EOMs for the corresponding scalar perturbations. Notably, the EOMs contain terms that are higher order in time derivatives, which implies that with an arbitrary choice of values of $c_{1},\cdots,c_{5}$, the action (\ref{action}) may propagate additional degree of freedom and some of them may suffer from the Ostrogradsky instability.
Therefore we need to find conditions for the parameters $c_{1},\cdots,c_{5}$ such that no term with higher time derivative are present.
Given the complexity of EOMs and the fact that the exact expressions are not crucial to our discussion below, in the following we only present terms that involve higher-order time derivatives.
\begin{itemize}
 \item In the EOM of $\phi$:
\begin{equation}
\text{EOM}(\phi) \supset -4(c_1+c_2+c_3+c_4+c_5)C''',
\end{equation}
\item In the EOM of $\psi$:
\begin{equation}
\text{EOM}(\psi) \supset 6(2c_3+c_5)C''',
\end{equation}
\item In the EOM of $C$:
\begin{equation}
\begin{split}
\text{EOM}(C)\supset &-(2c_1+3c_2+4c_3+3c_4+4c_5)\partial^i\partial_i D'''\\&-16(c_1+c_2+c_3+c_4+c_5)\mathcal{H}C'''+4(c_1+c_2+c_3+c_4+c_5)\phi'''\\&-6(2c_3+c_5)\psi'''-4(c_1+c_2+c_3+c_4+c_5)C'''',
\end{split}
\end{equation}
\item In the EOM of $D$:
\begin{equation}
\begin{split}
\text{EOM}(D) \supset & (2c_1+3c_2+4c_3+3c_4+4c_5)\partial^i\partial_i C'''-4(2c_1+c_2+c_4)\mathcal{H}\partial^i\partial_iD'''\\&-(2c_1+c_2+c_4)\partial^i\partial_i D''''.
\end{split}
\end{equation}
\end{itemize}
There are no higher-order time derivative terms in the EOM of perturbation $\delta\varphi$. 

To avoid the possible Ostrogradsky instability, the coefficients of the higher-order time derivative terms should vanish. 
Therefore, the parameters $c_1,\cdots,c_5$ must satisfy the following constraints
\begin{gather}
\label{constraints1}
2c_3+c_5=0,\\
\label{constraints2}
2c_1+c_2+c_4=0,\\
\label{constraints3}
c_1+c_2+c_3+c_4+c_5=0,\\
\label{constraints4}
2c_1+3c_2+4c_3+3c_4+4c_5=0.
\end{gather}
Note that the above four equations \eqref{constraints1}-\eqref{constraints4} are not independent. Solving these equations yields the following solutions for the parameters:
\begin{equation}\label{set}
c_1=\frac{1}{2}c_5,\ \ c_2=-c_4-c_5,\ \ c_3=-\frac{1}{2}c_5.
\end{equation}
In Ref. \cite{Runkla:2018xrv}, the authors obtain the same results by demanding the second derivatives of the non-metricity tensor in the action to be vanishing. Note that the parameter set \eqref{ps}, which makes $\mathbb{Q}$ to become the symmetric teleparallel equivalent GR Lagrangian, is a special case of \eqref{set}. Furthermore, by substituting the solutions \eqref{set} into Eqs. \eqref{C1}-\eqref{C3}, we find that $\mathcal{C}_1=-3c_5$, $\mathcal{C}_2=0$ and $\mathcal{C}_3=-2c_5$, which implies that there is no extra constraint imposed on the background equations \eqref{p14}.

In the rest of this work, we will perform the calculation with the solutions \eqref{set}, which can avoid the Ostrogradsky instability for the linear perturbations.

With Eq. \eqref{set}, the quadratic action for the scalar perturbations \eqref{L2} reduces to be
\begin{equation}\label{reduceA2}
\begin{split}
S^{(2)}_{SS}=&\int \mathrm{d}^3x \mathrm{d}\tau a^2\left[-\frac{1}{2}a^2V_{\varphi\varphi}\delta\varphi^2-a^2V_{\varphi}\delta\varphi\phi+3a^2V_{\varphi}\delta\varphi\psi-\frac{1}{2}\partial^i\delta\varphi\partial_i\delta\varphi+\frac{1}{2}(\delta\varphi')^2\right.\\&\left.
-(\phi+3\psi)\delta\varphi'\varphi'+4c_5\partial^i\psi\partial_i\phi-2c_5\partial^i\psi\partial_i\psi+6c_5\left(2\mathcal{H}\phi\psi'+6\mathcal{H}\psi\psi'+(\psi')^2\right)\right.\\&\left.+(9\psi^2+\phi^2)\left(6c_5\mathcal{H}^2+\frac{1}{2}(\varphi')^2\right)-4c_4\mathcal{H}\left(\partial_i\partial^iD-C'+\phi+\psi\right)\partial_i\partial^i\left(D'-C\right)\right],
\end{split}
\end{equation}
where we have used the background equations \eqref{p11}-\eqref{p13}. 
By varying the action \eqref{reduceA2} with respect to the scalar perturbations, we obtain the following EOMs for the linear scalar perturbations
\begin{gather}
\label{le1}
4c_4\mathcal{H}\partial_i\partial^i(C-D')+4c_5(3\mathcal{H}\psi'+3\mathcal{H}^2\phi-\partial_i\partial^i\psi)=-(\varphi')^2\phi+\delta\varphi'\varphi'+a^2V_{\varphi}\delta\varphi,\\
\label{le2}
4c_4\mathcal{H}\partial_i\partial^i(C-D')+4c_5\partial_i\partial^i(\psi-\phi)-12c_5\mathcal{H}'(\phi+3\psi)-12c_5\mathcal{H}(\phi'+2\psi')\\ \nonumber -12c_5\psi''+12c_5\mathcal{H}^2(3\psi-2\phi)+9(\varphi')^2\psi-3\delta\varphi'\varphi'+3a^2V_{\varphi}\delta\varphi=0,\\
\label{le3}
c_5(\psi-\phi)=c_4\mathcal{H}(D'-C),\\
\label{le4}
\delta\varphi''+2\mathcal{H}\delta\varphi'-\partial_i\partial^i\delta\varphi+a^2V_{\varphi\varphi}\delta\varphi+2a^2V_{\varphi}\phi-(\phi'+3\psi')\varphi'=0,\\
\label{le5}
\mathcal{H}\partial_i\partial^i(\phi+\psi-D''+\partial_i\partial^iD)+(2\mathcal{H}^2+\mathcal{H}')\partial_i\partial^i(C-D')=0 ~(c_4\neq 0),\\
\label{le6}
(2\mathcal{H}^2+\mathcal{H}')\partial_i\partial^i(\phi+\psi-C'+\partial_i\partial^i D)+\mathcal{H}\partial_i\partial^i(\phi'+\psi'-C''+\partial_i\partial^i C)=0 ~(c_4\neq 0).
\end{gather}

In Eq. \eqref{reduceA2}, the perturbations $C$ and $D$ disappear when $c_4=0$, which implies that they do not acquire the linear equations of motion of their own. However, in the next section, we will observe that $C$ and $D$ do exist at nonlinear orders and will contribute to the SIGWs, even in the case of $c_{4} = 0$. This is reminiscent of the so-called strong coupling problem in the study of the Ho\v{r}ava gravity and 
$f(T)$ gravity \cite{Blas:2009yd,Charmousis:2009tc,Blas:2009qj,Papazoglou:2009fj,Blas:2009ck,Ferraro:2018tpu,Li:2011rn,Izumi:2012qj,Hu:2023juh,Golovnev:2020zpv}, in which some perturbation modes do not show up at the linear order around a homogeneous and isotropic background, but do exist either at nonlinear orders or at linear order around an inhomogeneous background. Furthermore, from the quadratic action \eqref{reduceA2} we can see that if parameter $c_5=0$, the perturbation from the metric $\psi$ does not involve any time derivatives. As a result, $\psi$ becomes an auxiliary variable without dynamic behavior.

\subsection{The evolution of the background and the linear perturbation}\label{IIIC}
In order to calculate the SIGWs, in this subsection, we first discuss the evolution of the background and the linear scalar perturbations. 

Although we are left with only two independent parameters $c_4$ and $c_5$ under the condition \eqref{set} in order to remove higher-order time derivatives, the EOMs of background, \eqref{p11}-\eqref{p14}, and linear scalar perturbations, \eqref{le1}-\eqref{le6}, are still difficult to solve in general. 
Fortunately, we observe that Eqs. \eqref{le1}-\eqref{le6} simplify dramatically if $C=D'$, which by itself does not invalidate Eqs. \eqref{le1}-\eqref{le6}.
In other words,  $C=D'$ is a special solution for the Eqs. \eqref{le1}-\eqref{le6}. If we further choose the parameter $c_5=-1/2$, the scalar perturbations from the metric and the fluctuation of the scalar field are the same as those in GR. 
For our purpose to investigate the contributions of the non-metricity tensor and the PV term to the SIGWs, we expect that our model deviates from GR minimally, namely the evolution of background, the perturbations from metric, and the fluctuation of scalar field are the same as those in GR. As a result, we require $c_5=-1/2$, and $\phi=\psi$ without the presence of anisotropic stress. 
From Eq. \eqref{le3}, this is also consistent with the special solution $C=D'$.
We may view this choice of parameters and $C=D'$ as the minimal modification of GR in the framework of symmetric teleparallel gravity, which meanwhile evades the strong coupling problem. In the rest of this paper, we will evaluate the SIGWs with this minimal modification, while leaving $c_4$ as a free parameter.

During the radiation-dominated era, we have $\bar{P}/\bar{\rho}=1/3$, where
\begin{equation}
\bar{\rho}=\frac{1}{2a^2}(\varphi')^2+V,\ \bar{P}=\frac{1}{2a^2}(\varphi')^2-V.
\end{equation}
By making use of \eqref{set} together with the choice $c_5=-1/2$, the EOMs of the background, \eqref{p11}-\eqref{p14}, are the same as those in GR.
Thus we can obtain the evolution of the background during the radiation-dominated era \cite{Zhang:2022xmm},
\begin{equation}\label{beq}
\bar{\rho}=\rho_0 a^{-4},\ \ a=\sqrt{\frac{1}{3}\rho_0}\tau=a_0\tau, \ \ \varphi'=\pm2\tau^{-1}.
\end{equation}

With the solution $C=D'$ and $c_5=-1/2$, the EOMs of linear scalar perturbations from the metric and the fluctuation of the scalar field are also the same as those in GR. 
Additionally, with $C=D'$ the Eq. \eqref{le3} implies $\phi=\psi$.
As a result, the EOM of perturbation $D$ reduces to be
\begin{equation}\label{ED}
D''-\partial_i\partial^iD=2\phi.
\end{equation}

For late convenience of calculating the SIGWs, we split the perturbations into the primordial  perturbation and the transfer functions as follows,
\begin{gather}\label{trans1}
\phi(\bm k,\tau)=\frac{2}{3}\zeta(\bm k)T_\phi(x),\\
\label{trans2}
D(\bm k,\tau)=\frac{2}{3}\zeta(\bm k)\frac{1}{k^2}T_D(x),
\end{gather}
where $\zeta$ is the primordial curvature perturbation and $x=k\tau$. The transfer function $T_{\phi}$ is solved to be
\begin{equation}\label{Tphi}
T_{\phi}(x)=\frac{9}{x^2}\left(\frac{\sin(x/\sqrt{3})}{x/\sqrt{3}}-\cos(x/\sqrt{3})\right).
\end{equation}
Substituting the transfer function into Eq. \eqref{ED}, we can obtain the transfer function of $D$,
\begin{equation}\label{TD}
\begin{split}
T_{D}(x)=&C_1\cos(x)+C_2\sin(x)+\frac{9\sqrt{3}}{x}\sin(\frac{x}{\sqrt{3}})-3\sqrt{3}\cos(x)\left(\text{Ci}(x+\frac{x}{\sqrt{3}})-\text{Ci}(x-\frac{x}{\sqrt{3}})\right)\\&-3\sqrt{3}\sin(x)\left(\text{Si}(x+\frac{x}{\sqrt{3}})-\text{Si}(x-\frac{x}{\sqrt{3}})\right),
\end{split}
\end{equation}
where
\begin{equation}
\text{Si}(x)=\int_0^x \mathrm{d} y\frac{\sin y}{y},\quad \text{Ci}(x)=-\int_x^\infty \mathrm{d} y \frac{\cos y}{y}
\end{equation}
are sine integral and cosine integral, respectively.
In Eq. \eqref{TD}, $C_1, C_2$ are integral constants. We expect the perturbation decays and tends to $0$ during the radiation-dominated era, therefore we choose $C_1=C_2=0$. 

\section{The scalar induced gravitational waves}\label{sec4}

In this section, we derive the EOM for the SIGWs. 
To this end, we expand the action \eqref{action} up to the third order and focus on terms that are quadratic in the scalar perturbation and linear in the tensor perturbations, which correspond to the source term in the EOM for the SIGWs. 
The action that is relevant to the SIGWs is given by
\begin{equation}\label{acgws}
S_{\mathrm{GW}}=S^{(2)}_{TT}+S^{(3)}_{SST},
\end{equation}
where
\begin{equation}
S^{(2)}_{TT}=\int \mathrm{d}^3x \mathrm{d}\tau a^2\left[\frac{1}{8}\left(h^{'}_{ij}h^{'ij}-\partial_k h_{ij}\partial^k h^{ij}\right)+\frac{1}{2}\mathcal{M}\epsilon^{ijk}\partial_j h_{kl}h^{\ l}_{i}\right],
\end{equation}
is the quadratic action for the tensor perturbations with
\begin{equation}\label{MM}
\mathcal{M}=2(2\mathcal{H}g(\varphi)+g'(\varphi)).
\end{equation}
The cubic action involving two scalar modes and one the tensor modes is 
\begin{equation}
\begin{split}
S^{(3)}_{SST}=\int \mathrm{d}^3x \mathrm{d}\tau a^2\left(\mathcal{L}^{\mathrm{PC}}_{ij}+\mathcal{L}^{\mathrm{PV}}_{ij}\right)h^{ij},
\end{split}
\end{equation}
where
\begin{equation}\label{L1}
\begin{split}
\mathcal{L}^{\mathrm{PC}}_{ij}= &\frac{1}{2}\partial_i\delta\varphi\partial_j\delta\varphi-2c_5\partial_i\phi\partial_j\psi-c_4\left(2\mathcal{H}\phi\partial_i\partial_j D'+\phi'\partial_i\partial_j D'+\phi\partial_i\partial_j D''\right.\\&\left.+2\partial^k\phi\partial_k\partial_j\partial_i D+2\phi\partial^k\partial_k\partial_i\partial_j D+\partial_j C'\partial_i\phi-\phi'\partial_i\partial_j C+2\mathcal{H}\phi\partial_i\partial_jC-3\psi'\partial_i\partial_jC\right.\\&\left.+2\mathcal{H}\psi\partial_i\partial_j C-3\psi\partial_i\partial_j C'+2\mathcal{H}\psi\partial_i\partial_j D'+3\psi'\partial_i\partial_j D'+3\psi\partial_i\partial_j D''-2\partial^k\psi\partial_k\partial_i\partial_j D\right.\\&\left.-2\psi\partial^k\partial_k\partial_i\partial_j D-\partial_i\partial_j C\partial^k\partial_k C+\partial_k\partial_i C\partial^k\partial_j C+C'\partial_i\partial_j C'+C''\partial_i\partial_j C-2\mathcal{H}C'\partial_i\partial_j C\right.\\&\left.+2\mathcal{H}\partial_i C'\partial_j D'+\partial_i D'\partial_jC''+\partial_i C'\partial_jD''-2\mathcal{H}\partial^k C\partial_k\partial_i\partial_j D+\partial_i C'\partial^k\partial_k\partial_j D\right.\\&\left.-2\partial_i\partial_k D'\partial^k\partial_j C-\partial_i D'\partial^k\partial_k\partial_j C+2\mathcal{H}\partial^k\partial_j D'\partial_k\partial_i D+2\partial^k\partial_jD''\partial_k\partial_i D\right.\\&\left.+\partial_k\partial_j D'\partial^k\partial_i D'+\partial_iD''\partial^k\partial_k\partial_j D+2\partial^k\partial_i\partial_j D\partial^l\partial_l\partial_k D-2\partial_l\partial_k\partial_j D\partial^l\partial^k\partial_i D\right)
\end{split}
\end{equation}
correspond to terms that preserve the parity symmetry, 
and
\begin{equation}\label{LPV}
\begin{split}
\mathcal{L}^{\mathrm{PV}}_{ij}= &\mathcal{M}\epsilon_{jkl}(\partial^k\partial^m D\partial^l\partial_m\partial_i D-\partial^k\partial_i C\partial^l D')+2g_{\varphi}\epsilon_{jkl}\partial^l\delta\varphi(\partial^k\partial_i C+\partial^k\partial_i D')\\&+2g\epsilon_{jkl}(2\partial^k\partial_i C\partial^l\phi-2\partial^k\partial_i D'\partial^l\phi+4\partial^k\psi\partial^l\partial_i D'\\&-\partial^k\partial_i C\partial^l C'-\partial^k C'\partial^l\partial_i D'-\partial^k\partial_i C\partial^l D''-2\partial^k\partial_i\partial_m D\partial^l\partial^m C\\&+2\partial^k\partial^m D\partial^l\partial_m\partial_i D'-\partial^k D''\partial^l\partial_i D')
\end{split}
\end{equation}
correspond to terms that are parity-violating, respectively.

By varying the action \eqref{acgws} with respect to the tensor perturbations $h^{ij}$, we obtain the EOM for the SIGWs,
\begin{equation}\label{EOMGWs}
-\frac{1}{4}\left(h^{''}_{ij}+2\mathcal{H}h^{'}_{ij}-\nabla^2h_{ij}\right)+\frac{1}{2}\mathcal{M}\left(\epsilon_{ilk}\partial_l h_{kj}+\epsilon_{jlk}\partial_l h_{ki}\right)=\mathcal{T}^{lm}_{\ \ ij}s_{lm},
\end{equation}
where $\mathcal{T}^{lm}_{\ \ ij}$ is the projection tensor, and the source reads
\begin{equation}
s_{ij}=-\frac{1}{2}(\mathcal{L}^{\mathrm{PC}}_{ij}+\mathcal{L}^{\mathrm{PC}}_{ji}+\mathcal{L}^{\mathrm{PV}}_{ij}+\mathcal{L}^{\mathrm{PV}}_{ji}).
\end{equation}
In the above, we have symmetrized the source with respect to $i\leftrightarrow j$.

According to Eq. \eqref{reduceA2}, if $c_4=0$, the perturbations $C$ and $D$ drop out in the quadratic action \eqref{reduceA2}.
However, they do appear in the cubic action of the SIGWs \eqref{acgws} even in the case of $c_{4} = 0$, which results in the strong coupling problem. 

In order to solve the EOM of SIGWs \eqref{EOMGWs}, we decompose $h_{ij}$ into circularly polarized modes as
\begin{equation}
h_{ij}(\bm{x},\tau)=\sum\limits_{A=R,L}\int \frac{\mathrm{d}^3k}{(2\pi)^{3/2}}e^{i\bm{k}\cdot\bm{x}}p^{A}_{ij}h^A_{\bm{k}}(\tau),
\end{equation}
where the circular polarization tensors are defined as
\begin{equation}\label{cpt}
p^R_{ij}=\frac{1}{\sqrt{2}}(\mathbf e^{+}_{ij}+i\mathbf e^{\times}_{ij}), \ \ p^L_{ij}=\frac{1}{\sqrt{2}}(\mathbf e^{+}_{ij}-i\mathbf e^{\times}_{ij}).
\end{equation}
The plus and cross polarization tensors can be expressed as
\begin{equation}
\label{poltensor1}
\begin{split}
\mathbf e^+_{ij}=&\frac{1}{\sqrt{2}}(\mathbf e_i \mathbf e_j-\bar{\mathbf e}_i \bar{\mathbf e}_j),\\
\mathbf e_{ij}^\times=&\frac{1}{\sqrt{2}}(\mathbf e_i\bar{\mathbf e}_j+\bar{\mathbf e}_i \mathbf e_j),
\end{split}
\end{equation}
where ${\mathbf e_{i}\left(\bm{k}\right)}$ and ${\bar{\mathbf e}_{i}\left(\bm{k}\right)}$ are two basis vectors which are orthogonal to each other and perpendicular to the wave vector ${\bm{k}}$, i.e., satisfying ${\bm k}\cdot {\mathbf e}={\bm k}\cdot \bar{\mathbf e}={\mathbf e}\cdot \bar{\mathbf e}=0$ and $|{\mathbf e}|=|\bar{\mathbf e}|=1$.

In Eq. \eqref{EOMGWs}, the projection tensor  extracts the transverse and trace-free part of the source, of which the definition is
\begin{equation}
\mathcal{T}^{lm}_{\ \ \ ij}s_{lm}(\bm{x},\tau)=\sum\limits_{A=R,L}\int \frac{\mathrm{d}^3\bm k}{(2\pi)^{3/2}}e^{i {\bm k} \cdot {\bm x}}p_{ij}^A p^{Alm}\tilde{s}_{lm}(\bm k,\tau), 
\end{equation}
where $\tilde{s}_{ij}$ is the Fourier transformation of the source $s_{ij}$.

With the above settings, we can now rewrite the EOM of SIGWs in Fourier space as
\begin{equation}\label{eu}
u^{A''}_{\bm k}+\left(\omega^2_A-\frac{a''}{a}\right)u^{A}_{\bm k}=-4aS^A_{\bm{k}},
\end{equation}
where  $u^A=ah^A$,
	\begin{equation}\label{omgA}
		\omega^2_A=k^2-4\mathcal{M}\lambda^Ak, \ \ (\lambda^R=1,\ \lambda^L=-1),
	\end{equation}
and
\begin{equation}
S^A_{\bm{k}}=p^{Aij}\tilde{s}_{ij}(\bm{k},\tau).
\end{equation}

The source $S^A_{\bm k}$ can be divided into two parts: the parity-conserved  part and the parity-violating part, given by
\begin{equation}
S^A_{\bm k}=S^{A(\mathrm{PC})}_{\bm k}+S^{A(\mathrm{PV})}_{\bm k},
\end{equation}
where 
\begin{equation}
\begin{split}
 S^{A(\mathrm{PC})}_{\bm k}=&\int \frac{\mathrm{d}^3\bm k'}{(2\pi)^{3/2}}p^{Aij}k^{'}_i k^{'}_j\zeta(\bm k')\zeta(\bm k-\bm k') f_{\mathrm{PC}}(u,v,x),\\
 S^{A(\mathrm{PV})}_{\bm k}=&\int \frac{\mathrm{d}^3\bm k'}{(2\pi)^{3/2}}p^{Aij}k^{'}_i k^{'}_j\zeta(\bm k')\zeta(\bm k-\bm k') f_{\mathrm{PV}}(k,u,v,x),
\end{split}
\end{equation}
and  $u=k'/k$, $v=|\bm k-\bm{k}'|/k$.
Note
\begin{equation}
p^{Aij}k^{'}_i k^{'}_j=\frac{1}{2}k^{'2}\sin^2(\theta)\text{e}^{2i\lambda^A\ell},
\end{equation}
where $\theta$ is the angle between $\bm{k}'$ and $\bm{k}$ while $\ell$ is the azimuthal angle of $\bm{k}'$. The function $f_{\mathrm{PC}}(u,v,x)$ and $f^A_{\mathrm{PV}}(u,v,x)$ are defined as
\begin{equation}\label{fPC}
\begin{split}
f_{\mathrm{PC}}(u,v,x)=&-\frac{2}{9}\left[\frac{1}{2}\frac{(ukT^{*}_{\psi}(ux)+\mathcal{H}T_{\phi}(ux))(vkT^{*}_{\psi}(vx)+\mathcal{H}T_{\phi}(vx))}{\mathcal{H}^2-\mathcal{H}'}+T_{\phi}(ux)T_{\phi}(vx)\right.\\&\left.-c_4\left(-\frac{8\mathcal{H}}{vk}T_{\phi}(ux)T^{*}_{D}(vx)+\frac{4\mathcal{H}}{vk}T^{**}_{D}(ux)T^{*}_{D}(vx)\right.\right.\\&\left.\left.-2\mathcal{H}\frac{1-u^2-v^2}{uv^2k}T^{*}_{D}(ux)T_{D}(vx)-\frac{1-u^2+v^2}{v^2}T^{**}_{D}(ux)T_{D}(vx)\right.\right.\\&\left.\left.-\frac{(1-u^2-v^2)(1-u^2+v^2)}{2u^2v^2}T_{D}(ux)T_{D}(vx)\right)+(u\leftrightarrow v)\right],
\end{split}
\end{equation}
and
\begin{equation}\label{fPV}
\begin{split}
f^A_{\mathrm{PV}}(u,v,x)= &-\frac{2}{9}\lambda^A\left[\mathcal{M}\left(\frac{1-u^2-v^2}{2uv^2k}T_D(ux)T_D(vx)-\frac{1}{vk}T^{*}_D(ux)T^{*}_D(vx)\right)\right.\\&\left.+4g_{\varphi}\varphi'\frac{u}{v}\frac{ukT^{*}_{\phi}(ux)+\mathcal{H}T_{\phi}(ux)}{(\mathcal{H}^2-\mathcal{H'})}T^{*}_D(vx)\right.\\&\left.+2g\left(-\frac{4u}{v}T_{\psi}(ux)T^{*}_{D}(vx)+ 2\frac{u-v}{v}T^{**}_{D}(ux)T^{*}_D(vx)\right.\right.\\&\left.\left.+2\frac{1-u^2-v^2}{uv}T_D(ux)T^{*}_{D}(vx)\right)+u\leftrightarrow v\right],
\end{split}
\end{equation}
respectively. The $\ast$ represents  derivatives with respect to the arguments. In deriving Eqs. \eqref{fPC} and \eqref{fPV}, we have used the relations $C=D'$, $\phi=\psi$, and
\begin{equation}
\delta\varphi=\frac{\psi'+\mathcal{H}\phi}{\mathcal{H}^2-\mathcal{H}'}\varphi'.
\end{equation}

Eq. \eqref{eu} can be solved by the method of Green's function,
\begin{equation}\label{Eh}
h^{A}_{\bm k}\left(\tau\right)=-\frac{4}{a(\tau)}\int^{\tau}\mathrm{d}\bar{\tau}~G^A_{k}\left(\tau,\bar{\tau}\right)
a\left(\bar{\tau}\right)S^A_{\bm k}\left(\bar{\tau}\right),
\end{equation}
where the Green's function $G^A_{k}\left(\tau,\bar{\tau}\right)$ satisfies the equation
\begin{equation}\label{GREEN}
G^{A''}_{k}(\tau,\bar{\tau})+\left(\omega_A^2-\frac{a''}{a}\right)G^{A}_{k}(\tau,\bar{\tau})=\delta(\tau-\bar{\tau}).
\end{equation}
As for the Green's function, the deviation from the standard GR is characterized by the parameter $\mathcal{M}$.
Generally, since $\omega_A$ given in Eq. (\ref{omgA}) is an involved function of both the wave number $k$ and the conformal time $\tau$ (see Eq. (\ref{MM})), it is difficult to solve Eq. \eqref{GREEN} and get the expression for the Green's function analytically. Nevertheless, for our purpose of studying the contributions of the scalar perturbations to the SIGWs, we assume the change of the Green's function from that in GR is also ``minimally''. Precisely, since $\omega_A$ is related to the propagating speeds of the GWs, we assume that in the duration of generation of SIGWs, $\omega_A$ is approximately time independent and depends only on the wave number. 
In fact, an exponential form of the coupling function
\begin{equation}\label{gvp}
g(\varphi)=g_0\mathrm{e}^{\alpha\varphi},
\end{equation}
renders $\omega_A$ independent of  time and allows us to obtain an analytical solution of Eq. \eqref{GREEN}.

Using the background Eqs. \eqref{beq}, the solution of the scalar field is found to be
\begin{equation}\label{evp}
\varphi=2\beta\ln(\tau/\tau_0)+\varphi_0,
\end{equation}
where $\varphi_0$ is the value of $\varphi$ at $\tau_0$ and $\beta=\pm1$, which corresponds to $\varphi'=\pm 2/\tau$, respectively.
Substituting Eqs. \eqref{gvp} and \eqref{evp} into the definition of $\mathcal{M}$, we have
\begin{equation}
\mathcal{M}=\frac{4(1+\alpha\beta)g_0\mathrm{e}^{\alpha\varphi_0}\tau^{2\alpha\beta-1}}{\tau^{2\alpha\beta}_0}. \label{Mgen}
\end{equation}
From Eq. (\ref{Mgen}), it is clear that if we set $2\alpha\beta-1=0$, $\mathcal{M}$ becomes constant.
As a result,
\begin{equation}
\omega^2_A=k^2\left(1-\frac{4\lambda^A\mathcal{M}_0}{k}\right),
\end{equation}
with 
	\begin{equation}
		\mathcal{M}_0=6g_0\mathrm{e}^{\alpha\varphi_0}/\tau_0, \label{calM0}
	\end{equation} 
which is independent of time. 
With these assumptions, we can solve Eq. \eqref{GREEN} analytically to get the expression of Green's function,
\begin{equation}
G^{A}_{k}(\tau,\bar\tau)=\frac{\sin[\omega_A(\tau-\bar\tau)]}{\omega_A}\Theta(\tau-\bar\tau), \label{gf}
\end{equation}
where $\Theta$ is the Heaviside step function.

The constant $\mathcal{M}_{0}$ defined in Eq. (\ref{calM0}) has the dimension of energy, which can be viewed as the characteristic energy scale of parity violation in our model.
It is therefore interesting to have an estimation of $\mathcal{M}_{0}$ based on the current observation.
The recent observations from GW170817 \cite{LIGOScientific:2017vwq} and GRB170817A \cite{LIGOScientific:2017zic} constrain the speed of GWs to be
\begin{equation}
-3\times 10^{-15}\leq c_{\rm {gw}}-1\leq 7\times 10^{-16}.
\end{equation}
Recalling the definition of $\omega_A$ in Eq. (\ref{omgA}), 
\begin{equation}
c_{\rm {gw}}=\frac{\omega_A}{k}=\left(1-\frac{4\mathcal{M}_0\lambda^A}{k}\right)^{1/2}\simeq 1-\frac{2\mathcal{M}_0\lambda^A}{k},
\end{equation}
which means
\begin{equation}
\frac{|\mathcal{M}_0|}{k}<3.5\times 10^{-16}.
\end{equation}
Therefore, the typical energy scale of parity violation is much smaller than the wave numbers of interest.

In Ref. \cite{Wu:2021ndf}, the authors constrain $\mathcal{M}_0$ with  the GW events of binary black hole merger (BBH) in the LIGO-Virgo catalogs GWTC-1 and GWTC-2, the result is $\mathcal{M}_0<1.6 \times 10^{-42}\text{Gev}\sim \mathcal{O}(10^{-3})\ \text{Mpc}^{-1}$. Since the SIGWs generate on small scales, $k\gg1 \ \text{Mpc}^{-1}$, we have $\mathcal{M}_0/k\ll 1$. From the EOM of SIGWs \eqref{EOMGWs} and the source term \eqref{fPV}, the PV term is also suppressed by $|\mathcal{M}_0|/k$, namely, $f^A_{\mathrm{PV}}\propto \mathcal{M}_0/k$, which means the effect of PV term on SIGWs is negligible.

\section{The power spectra of the SIGWs}\label{sec5}

The solutions of the circularly polarized modes can be written in a compact form
\begin{equation}
\label{hsolution}
h^A_{\bm k}(\tau)=4 \int\frac{\mathrm{d}^3\bm k'}{(2\pi)^{3/2}} p^{Aij}k^{'}_i k^{'}_j\zeta(\bm k')\zeta(\bm k-\bm k')\frac{1}{k^2}I^{A}(k,u,v,x),
\end{equation}
where
\begin{equation}
\begin{split}
\label{I_int}
I^{A}(k,u,v,x)&=-\int_0^x\mathrm{d}\bar{x}\frac{a(\bar{\tau})}{a(\tau)}k G^A_{k}(\tau,\bar{\tau})\left(f_{\mathrm{PC}}(u,v,\bar x)+f^A_{\mathrm{PV}}(u,v,\bar x)\right)\\&
= I^{A}_{\mathrm{PC}}(k,u,v,x)+I^{A}_{\mathrm{PV}}(k,u,v,x),
\end{split}
\end{equation}
with
\begin{equation}
\label{ISC}
I^{A}_{\mathrm{PC}}(k,u,v,x)=-\int_0^x\mathrm{d}\bar{x}\frac{a(\bar{\tau})}{a(\tau)}kG^A_{k}(\tau,\bar{\tau})f_{\mathrm{PC}}(u,v,\bar x),\\
\end{equation}
and
\begin{equation}
\label{IPV}
I^{A}_{\mathrm{PV}}(k,u,v,x)=-\int_0^x\mathrm{d}\bar{x}\frac{a(\bar{\tau})}{a(\tau)}k G^A_{k}(\tau,\bar{\tau})f^A_{\mathrm{PV}}(u,v,\bar x).
\end{equation}
According to whether the perturbations $C$ and $D$ contribute or not, we can split $I^A_{\mathrm{PC}}$ into two parts, which we denote $I^A_{\mathrm{PC}1}$ and $I^A_{\mathrm{PC}2}$, respectively. 
Specifically, $I^A_{\mathrm{PC}1}$ does not include contributions from $C$ and $D$, which correspond to the first two terms of $f_{\mathrm{PC}}$. The analytic expression for $I^A_{\mathrm{PC}1}$ can be found in Appendix \ref{kernel}. The other parts, $I^A_{\mathrm{PC}2}$ and $I^A_{\mathrm{PV}}$ cannot be calculated analytically, so we will compute them numerically.

The power spectra of the SIGWs $\mathcal{P}_{h}^{A}$ are defined by 
\begin{equation}
\langle h^A_{\bm{k}} h^C_{\bm{k}'}\rangle =\frac{2\pi^2}{k^3}\delta^3(\bm k+\bm k')\delta^{AC}\mathcal{P}^{A}_{h}(k).
\end{equation}
With the above definition of $\mathcal{P}_{h}^{A}$ and the solution of SIGWs, we can obtain the power spectra of the SIGWs \footnote{Here, we assume $\zeta$ is Gaussian, please refer to Refs. \cite{Cai:2018dig,Unal:2018yaa,Adshead:2021hnm,Garcia-Saenz:2022tzu,Garcia-Saenz:2023zue} and references therein for the non-Gaussian effects.}
\begin{align}
\label{PStensor}
\mathcal{P}^{A}_h(k,x)=4\int_{0}^\infty\mathrm{d}u\int_{|1-u|}^{1+u}\mathrm{d}v
\mathcal{J}(u,v)I^{A}(u,v,x)^2\mathcal{P}_\zeta(uk)\mathcal{P}_\zeta(vk),
\end{align}
where
\begin{equation}
\mathcal{J}(u,v)=\left[\frac{4u^2-(1+u^2-v^2)^2}{4uv}\right]^2,
\end{equation}
and $\mathcal{P}_\zeta$ is the power spectrum of primordial curvature perturbation.

The fractional energy density of the SIGWs is \footnote{Note that in the literature \cite{Kohri:2018awv}, there is an additional $1/2$ in front of $h_{ij}$ defined in metric, and the prefactor of $\Omega_{\mathrm{GW}}$ is $1/48$. Of course, the results are independent of the definition of $h_{ij}$ \cite{Garcia-Saenz:2023zue}.}
\begin{equation}\label{OGW}
\begin{split}
\Omega_{\mathrm{GW}}(k,x)&=\frac{1}{12}\left(\frac{k}{\mathcal{H}}\right)^2\sum\limits_{A=R,L}\overline{\mathcal{P}^A_h(k,x)}=\frac{x^2}{12}\sum\limits_{A=R,L}\overline{\mathcal{P}^A_h(k,x)}\\
&=\frac{1}{3}\int_{0}^\infty\mathrm{d}u\int_{|1-u|}^{1+u}\mathrm{d}v
\mathcal{J}(u,v)\sum\limits_{A=R,L}\overline{\tilde{I}^{A}(k,u,v,x)^2}\mathcal{P}_\zeta(uk)\mathcal{P}_\zeta(vk),
\end{split}
\end{equation}
where the overline represents the time average, and $\overline{\tilde{I}^{A}(k,u,v,x)^2}=\overline{I^{A}(k,u,v,x)^2}x^2$. 
The GWs behave as free radiation, thus the fractional energy density of the SIGWs at the present time $\Omega_{\mathrm{GW},0}$ can be expressed as 
\cite{Espinosa:2018eve}
\begin{equation}\label{EGW}
\Omega_{\mathrm{GW},0}\left(k\right)=\Omega_{\mathrm{GW}}\left(k,\eta\rightarrow\infty\right)\Omega_{r,0},
\end{equation}
where $\Omega_{r,0}$ is the current fractional energy density of the radiation and approximately $9\times 10^{-5}$ \cite{Sato-Polito:2019hws}.

In order to analyze the features of the SIGWs in our model, we use a concrete power spectrum of the primordial curvature perturbation to compute the energy density of the SIGWs. Consider the energy density of SIGWs induced by the monochromatic power spectrum,
\begin{equation}\label{ps1}
\mathcal{P}_\zeta(k)=\mathcal{A}_\zeta\delta(\ln(k/k_p)),
\end{equation}
then we obtain the energy density of SIGWs at the present time
\begin{align}
\label{PStensor}
\Omega_{\text{GW},0}(k)=\frac{1}{3}\Omega_{r,0}\mathcal{A}_{\zeta}^2\tilde{k}^{-2}\mathcal{J}(\tilde{k}^{-1},\tilde{k}^{-1})\sum\limits_{A=R,L}\overline{\tilde{I}^{A}(k,\tilde{k}^{-1},\tilde{k}^{-1},x\rightarrow \infty)^2}\Theta(2-\tilde{k}),
\end{align}
where $\tilde{k}=k/k_p$.

\begin{figure}[htp]
\centering
\includegraphics[width=0.8\linewidth]{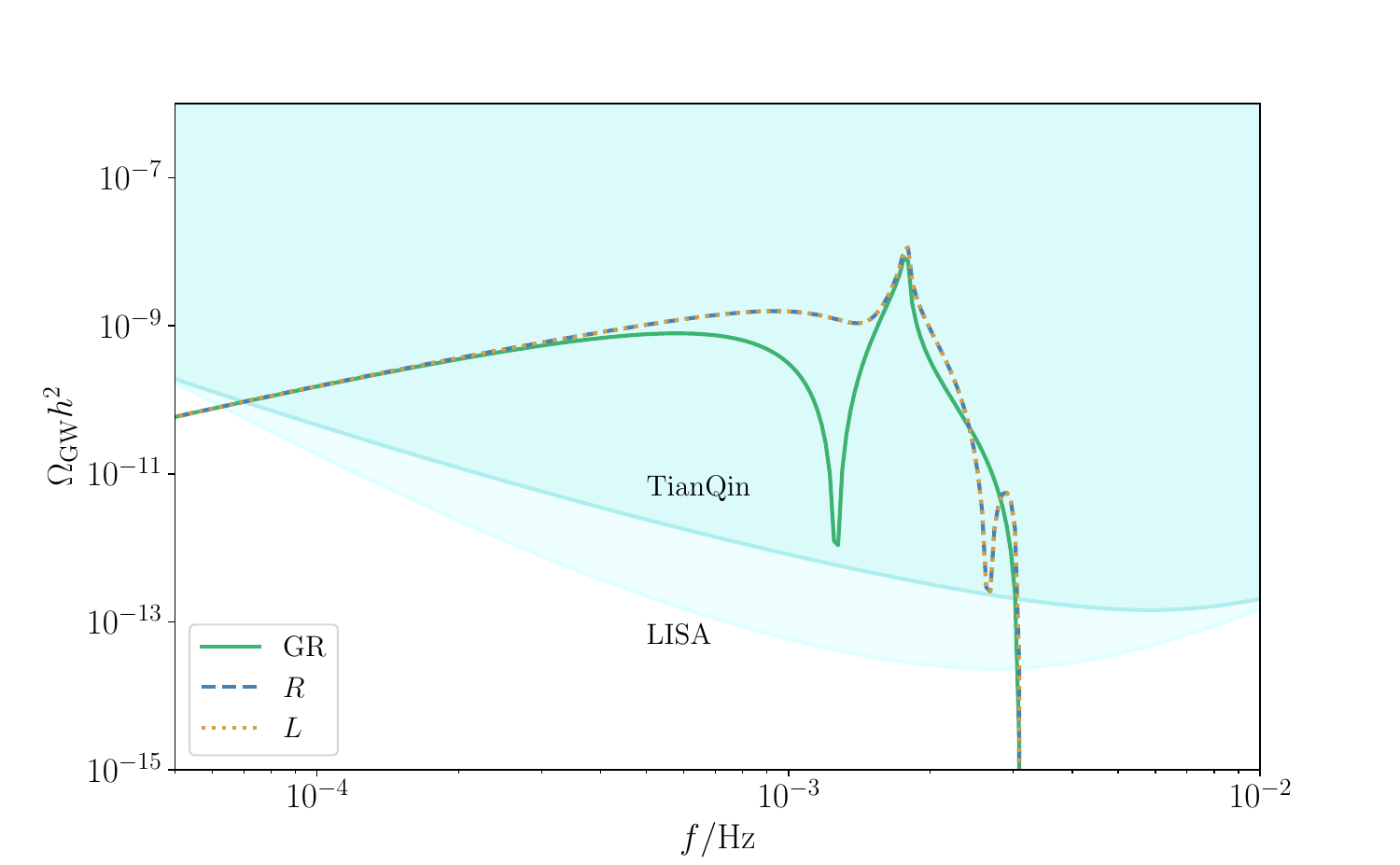}
\caption{The energy density of SIGWs induced by the monochromatic power spectrum. In this figure, we choose $c_4=-1/2$. The peak scale is $k_p=10^{12} \text{Mpc}^{-1}$, which corresponds to the maximum sensitivity of TianQin and LISA. The amplitude of the power spectrum is fixed to be $\mathcal{A}_{\zeta}=10^{-2}$.}
\label{fig:GWs1}
\end{figure}

\begin{figure}[htp]
\centering
\includegraphics[width=0.8\linewidth]{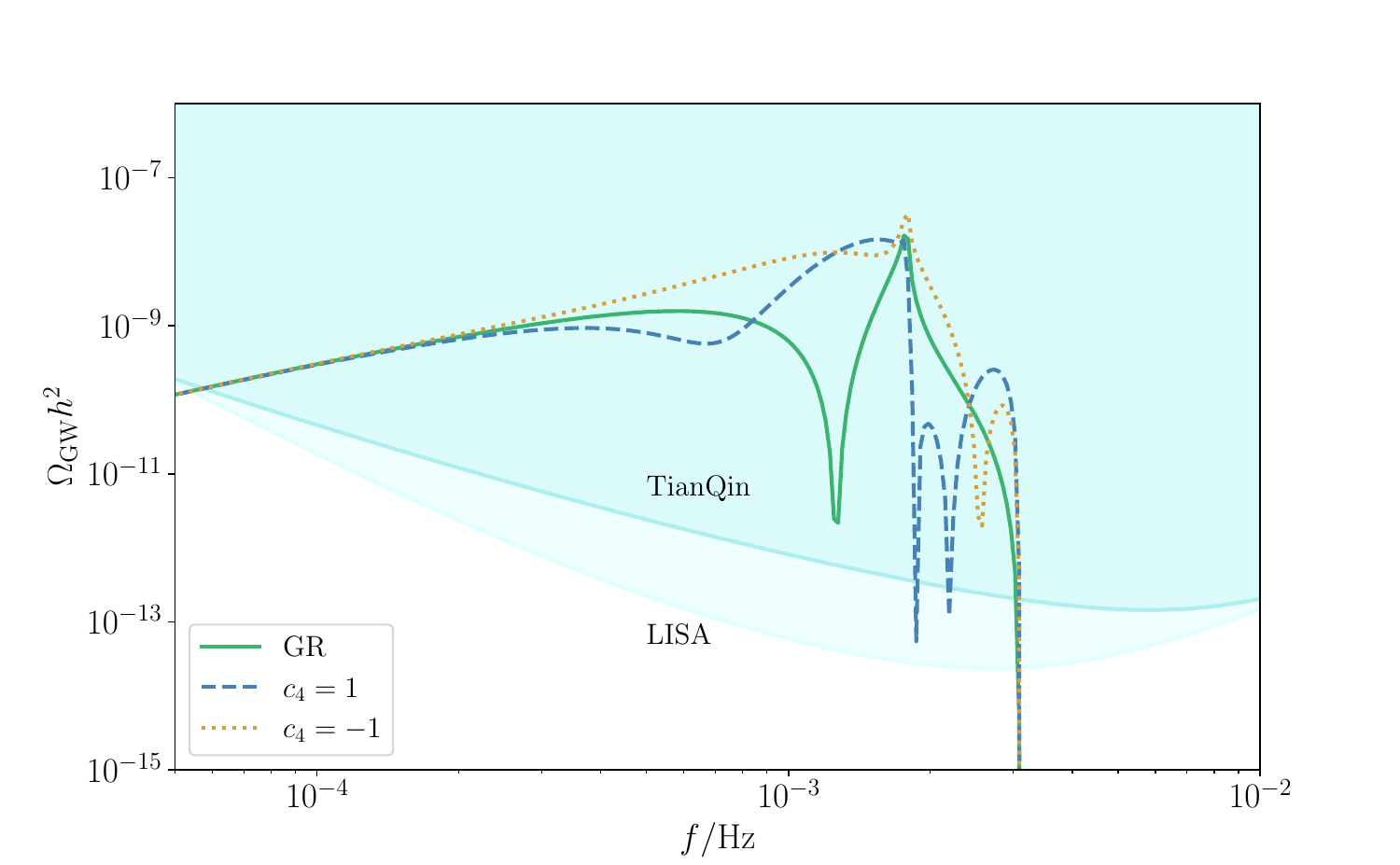}
\caption{The energy density of SIGWs induced by monochromatic power spectrum. The parameters of the monochromatic power spectrum are $k_p=10^{12} \text{Mpc}^{-1}$ and  $\mathcal{A}_{\zeta}=10^{-2}$.}
\label{fig:GWs2}
\end{figure}

We numerically calculate the energy density of SIGWs  and show the results in Figs. \ref{fig:GWs1} and \ref{fig:GWs2}. In order to compare the energy density of SIGWs in our model with that in GR, we also present the results of GR.
According to Fig. \ref{fig:GWs1}, the energy density of SIGWs from the left-hand polarized mode is almost the same as that from the right-hand polarized mode, which means that the effect from the PV term on the SIGWs is negligible. However,  the contributions from the perturbations $C$ and $D$ can have a significant impact on SIGWs, particularly at peak scales, 
which can be seen in both Figs. \ref{fig:GWs1} and \ref{fig:GWs2}.

The SIGWs in our model also exhibit some other interesting features. In GR, the scalar perturbation oscillates in the manner of $\sin(1/\sqrt{3}x)$ and $\cos(1/\sqrt{3}x)$, there is a divergence at $\tilde{k}=2/\sqrt{3}$ due to the resonant amplification \cite{Ananda:2006af,Kohri:2018awv}. In our model, the perturbations $C$ and $D$ behave differently, which have other oscillatory manners, namely of $\sin(x)$ and $\cos(x)$. This results in the resonant amplification at other scales. From Fig. \ref{fig:GWs2}, we can see that another peak appears at $\tilde{k}=1+1/\sqrt{3}$ in the case of the monochromatic power spectrum. This multipeak feature can be used to distinguish our model from GR.

\section{Conclusion}\label{seccon}

In this paper, we calculated the SIGWs in symmetric teleparallel gravity with a simple PV term $\widetilde{Q}Q$. The action of our model is given in Eq. (\ref{action}). 
In order to evade the strong coupling problem that has shown up in Ref. \cite{Li:2021mdp}, we replace the teleparallel equivalent Einstein-Hilbert action by a general non-metricity scalar $\mathbb{Q}$, which is a linear combination of scalar monomials that are quadratic in the non-metricity tensor. 
Under the requirement of no higher-order time derivative terms in the EOMs of linear scalar perturbations such that the possible Ostrogradsky instability is evaded, the constant parameters in $\mathbb{Q}$ must satisfy the constraint Eq. (\ref{set}) and only two parameters $c_4$ and $c_5$ are independent. The strong coupling problem can be avoided only when $c_4\neq 0$. 

We solved the EOMs of linear scalar perturbations and obtained their transfer functions during the radiation-dominated era. We have chosen the coupling function of the PV term to be the exponential form Eq. (\ref{gvp}), which ensures that the speed of SIGWs is independent of time.
We further derived the analytical expression for the Green's function of the tensor perturbations Eq. (\ref{gf}). 
We then calculated  the power spectra and the energy density of SIGWs. In order to analyze the features of SIGWs in our model, we evaluated numerically the energy density of SIGWs with a monochromatic power spectrum for the primordial curvature perturbation. 
Under the observation constraints on the propagating speeds of the GWs, we found that the effect of the PV term to the SIGWs is negligible. 
However, the contribution to SIGWs from the perturbations of connection can be significant, and results in a multipeak structure in the energy density of SIGWs.
This feature makes our model, and in fact more general symmetric teleparallel gravity theories, distinguishable from GR.

\begin{acknowledgments}
Fengge Zhang thanks Zheng Chen and Yang Yu for their helpful discussion. This work was supported by the National Natural Science Foundation of China (NSFC) under the grant No. 11975020 and No. 12005309. 
\end{acknowledgments}

\appendix

\section{The quadratic action of linear scalar perturbations}
\label{AC2}
The quadratic action of scalar perturbation is $S^{(2)}_{SS}=\int \mathrm{d}^3x \mathrm{d}\tau a^2 \mathcal{L}$, where
\begin{align}\label{L2}
\mathcal{L}=&\frac{1}{2}\left(\delta \varphi'\right)^2-\frac{1}{2}\partial_i \delta \varphi \partial^i \delta \varphi-\frac{1}{2} a^2 V_{\varphi \varphi} \delta \varphi^2-a^2 V_{\varphi} \delta \varphi \phi+3 a^2 V_{\varphi} \delta \varphi \psi \\ \nonumber & -(\phi+3\psi)\delta \varphi' \varphi'-2(c_1+c_2+c_3+c_4+c_5)\left(\phi'\right)^2-6(c_1+3c_3)\left(\psi'\right)^2\\ \nonumber& -2(c_1+c_2+c_3+c_4+c_5)\left(C''\right)^2\\ \nonumber&-\left((4c_1+c_2+16c_3+c_4+4c_5)\mathcal{H}^2-\frac{1}{4} \left(\varphi'\right)^2+\frac{1}{2} a^2 V\right) \phi^2\\ \nonumber& -6\left((4c_1+c_2+16c_3+c_4+4c_5)\mathcal{H}^2-\frac{1}{4} \left(\varphi'\right)^2-\frac{1}{2}a^2 V\right) \phi \psi\\ \nonumber&-9\left((4c_1+c_2+16c_3+c_4+4c_5)\mathcal{H}^2-\frac{1}{4} \left(\varphi'\right)^2+\frac{1}{2} a^2 V\right) \psi^2\\ \nonumber&+2(2c_1+2c_2+8c_3+2c_4+5c_5)\mathcal{H} \phi\phi'+6(2c_1+2c_2+8 c_3+2c_4+5c_5)\mathcal{H} \psi\phi'\\ \nonumber&-6(2c_1+8c_3+c_5)\mathcal{H} \phi\psi'-18(2c_1+8c_3+c_5)\mathcal{H} \psi \psi'+6(2c_3+c_5)\phi'\psi'\\ \nonumber&-2(2c_1+2c_2+8c_3+2c_4+5c_5)\mathcal{H} \phi C''-6(2c_1+2c_2+8c_3+2c_4+5c_5)\mathcal{H} \psi C'' \\ \nonumber&-2(2c_1+2c_2+8c_3+2c_4+5c_5)\mathcal{H} C' C''+4(c_1+c_2+c_3+c_4+c_5)\phi' C''\\ \nonumber&-6(2c_3+c_5)\psi' C''+2(3c_1+c_2+9c_3+c_4+3 c_5)\partial_i \psi \partial^i \psi+2(c_1+c_3)\partial_i \phi \partial^i \phi\\ \nonumber&-2(6c_3+c_5)\partial_i \psi \partial^i \phi-2(2c_1+c_2+12c_3+3c_4+6c_5)\mathcal{H} \phi \partial_i \partial^i D'\\ \nonumber
&-2(c_2+2c_3+c_4+2c_5)\phi \partial^i \partial_i D''\\ \nonumber&-6(2c_1+c_2+8c_3+c_4+3c_5)\mathcal{H} \psi\partial_i \partial^i D'-(4c_1+2c_2+12c_3+3c_5)\psi'\partial_i \partial^i D'\\ \nonumber
&-2(2c_3+c_5)\partial^i \phi \partial_j \partial^j \partial_i D+4(c_1+c_2+3c_3+c_4+2c_5)\partial^i \psi \partial_j \partial^j \partial_i D\\ \nonumber
&+(12c_3+2c_4+5c_5)\partial_i C' \partial^i \psi-(2c_4+3c_5)\partial_i D'' \partial^i \psi\\ \nonumber
&-(4c_1+2c_2+4c_3+c_5)\partial_i C' \partial^i \phi-2(c_2+c_4+2c_5)\mathcal{H} \phi \partial_i \partial^i C\\ \nonumber
&-(2c_4+c_5)\phi' \partial_i \partial^i C+2(c_2+c_4+2c_5)\mathcal{H} \psi \partial_i \partial^i C+(2c_2+3c_5)\psi' \partial_i \partial^i C\\ \nonumber
&+2(c_2+c_4+2c_5)\mathcal{H} C' \partial_i \partial^i C+\frac{1}{2}(6c_1+5c_2+4c_3+c_4+2c_5)\partial_i C' \partial^i C'\\ \nonumber
&-2(2c_1+c_2+8c_3+c_4+3c_5)\mathcal{H}\partial_i D'\partial^i C'\\ \nonumber
&-\frac{1}{2}(6c_1+5c_2+4c_3+c_4+2c_5)\partial_i\partial^i D' \partial_j\partial^j D'-(4c_3+2c_4+3c_5)\partial_i \partial^i D' C''\\ \nonumber
&-(2c_1+3c_2+c_4+c_5)\partial^i C' \partial_i D''+\frac{1}{2}(2c_1+c_2+c_4)\partial_i D'' \partial^i D''\\ \nonumber
&+(2c_1+3c_2+c_4+c_5)\partial_j \partial^j D' \partial_i \partial^i C+(2c_4+c_5) C'' \partial_i \partial^i C\\ \nonumber
&-2(2c_1+2c_2+8c_3+2c_4+5c_5)\mathcal{H}\partial_i D'' \partial^i C-\frac{1}{2}(2c_1+c_2+c_4)\partial_i \partial^i C \partial_j \partial^j C\\ \nonumber
&+(4c_3+2c_4+3c_5)\partial^i C' \partial_j \partial^j \partial_i D-(2c_4+c_5)\partial^i D'' \partial_j \partial^j \partial_i D\\ \nonumber
&+2(c_2+c_4+2c_5)\mathcal{H} \partial^i C \partial_j \partial^j \partial_i D-2(2c_1+c_2+8c_3+c_4+3c_5)\mathcal{H} \partial_j \partial_i D \partial^j \partial^i D'+\\ \nonumber
&+2(c_1+c_2+c_3+c_4+c_5)\partial_j \partial^j \partial^i D \partial_k \partial^k \partial_i D.
\end{align}

\section{The integral kernel}\label{kernel}
In this appendix, we give the integral kernel $I^A_{\text{PC}1}$. With the Green's function \eqref{GREEN}, $I^A_{\text{PC}1}$ can be expressed as 
\begin{equation}
I^A_{\text{PC}1}(k,u,v,x)=\frac{\sin(wx)}{wx}I^A_{\text{PC}1s}(k,u,v,x)+\frac{\cos(wx)}{wx}I^A_{\text{PC}1c}(k,u,v,x),
\end{equation}
where the subscript ``s'' and ``c'' stand for contributions involving the sine and cosine functions, respectively, and $w=\omega_A/k$. We also write
\begin{gather}
 I^A_{\text{PC}1s}(k,u,v,x)=\mathcal{I}^A_{\text{pc}1s}(k,u,v,x)- \mathcal{I}^A_{\text{pc}1s}(k,u,v,0), \nonumber\\
 I^A_{\text{PC}1c}(k,u,v,x)=\mathcal{I}^A_{\text{pc}1c}(k,u,v,x)- \mathcal{I}^A_{\text{pc}1c}(k,u,v,0),
\end{gather}
where $\mathcal{I}^A_{\text{pc}1s}$ and $\mathcal{I}^A_{\text{pc}1c}$ are defined by
\begin{gather}
\mathcal{I}^A_{\text{pc}1s}(k,u,v,y)=-\int \mathrm{d} y \cos wy f_{\text{PC}}(u,v,y) y, \nonumber\\
\mathcal{I}^A_{\text{pc}1c}(k,u,v,y)=\int \mathrm{d} y \sin wy f_{\text{PC}}(u,v,y) y.
\end{gather}
After lengthy calculations, we obtain
\begin{equation}
\begin{split}
\mathcal{I}^A_{\mathrm{pc}1s}(k,u,v,y)=&\frac{3}{2u^3v^3y^4}\left(-18uvy^2\cos\frac{uy}{\sqrt{3}}\cos\frac{v y}{\sqrt{3}}\cos wy +6uvwy^3\cos\frac{uy}{\sqrt{3}}\cos\frac{v y}{\sqrt{3}}\sin wy\right.\\&\left.-6\sqrt{3}vwy^2\cos\frac{vy}{\sqrt{3}}\sin\frac{u y}{\sqrt{3}}\sin wy-6\sqrt{3}uwy^2\cos\frac{uy}{\sqrt{3}}\sin\frac{v y}{\sqrt{3}}\sin wy\right.\\&\left.+\sqrt{3}vy(18-u^2y^2+v^2y^2-3w^2y^2)\cos\frac{vy}{\sqrt{3}}\cos wy\sin\frac{u y}{\sqrt{3}}\right.\\&\left.+\sqrt{3}uy(18-v^2y^2+u^2y^2-3w^2y^2)\cos\frac{uy}{\sqrt{3}}\cos wy\sin\frac{v y}{\sqrt{3}}\right.\\&\left.+3(18-u^2y^2-v^2y^2-3w^2y^2)\cos wy\sin\frac{uy}{\sqrt{3}}\sin\frac{v y}{\sqrt{3}}\right.\\&\left.+3wy(6+u^2y^2+v^2y^2-3w^2y^2)\sin\frac{uy}{\sqrt{3}}\sin\frac{v y}{\sqrt{3}}\sin wy\right)\\&-\frac{3(u^2+v^2-3w^2)^2}{8u^3v^3}\left(\text{Ci}\left[\left(w+\frac{u+v}{\sqrt{3}}\right)y\right]+\text{Ci}\left[\left|w-\frac{u+v}{\sqrt{3}}\right|y\right]\right.\nonumber\\&\left.\right.\\&\left.-\text{Ci}\left[\left(w+\frac{u-v}{\sqrt{3}}\right)y\right]-\text{Ci}\left[\left(w-\frac{u-v}{\sqrt{3}}\right)y\right]\right),
\end{split}
\end{equation}
and
\begin{equation}
\begin{split}
\mathcal{I}^A_{\mathrm{pc}1c}(k,u,v,y)=&\frac{3}{2u^3v^3y^4}\left(6uvwy^3\cos\frac{uy}{\sqrt{3}}\cos\frac{v y}{\sqrt{3}}\cos wy -6\sqrt{3}vwy^2\cos\frac{vy}{\sqrt{3}}\cos wy\sin\frac{u y}{\sqrt{3}}\right.\\&\left.-6\sqrt{3}uwy^2\cos\frac{uy}{\sqrt{3}}\cos wy\sin\frac{v y}{\sqrt{3}}+18uvy^2\cos\frac{uy}{\sqrt{3}}\cos\frac{v y}{\sqrt{3}}\sin wy\right.\\&\left.-\sqrt{3}vy(18-u^2y^2+v^2y^2-3w^2y^2)\cos\frac{vy}{\sqrt{3}}\sin\frac{u y}{\sqrt{3}}\sin wy\right.\\&\left.-\sqrt{3}uy(18-v^2y^2+u^2y^2-3w^2y^2)\cos\frac{uy}{\sqrt{3}}\sin\frac{v y}{\sqrt{3}}\sin wy\right.\\&\left.+3(18-u^2y^2-v^2y^2-3w^2y^2)\sin\frac{uy}{\sqrt{3}}\sin\frac{v y}{\sqrt{3}}\sin wy\right.\\&\left.+3wy(6+u^2y^2+v^2y^2-3w^2y^2)\cos wy\sin\frac{uy}{\sqrt{3}}\sin\frac{v y}{\sqrt{3}}\right)\\&+\frac{3(u^2+v^2-3w^2)^2}{8u^3v^3}\left(\text{Si}\left[\left(w+\frac{u+v}{\sqrt{3}}\right)y\right]+\text{Si}\left[\left(w-\frac{u+v}{\sqrt{3}}\right)y\right]\right.\nonumber\\&\left.\right.\\&\left.-\text{Si}\left[\left(w+\frac{u-v}{\sqrt{3}}\right)y\right]-\text{Si}\left[\left(w-\frac{u-v}{\sqrt{3}}\right)y\right]\right).
\end{split}
\end{equation}

We also have the following limits for $\mathcal{I}^A_{\mathrm{pc}1s}$
\begin{equation}
\mathcal{I}^A_{\mathrm{pc}1s}(u,v,y\rightarrow 0)=\frac{3(u^2+v^2-3w^2)}{8u^3v^3}\left(4uv-(u^2+v^2-3w^2)\log\left|\frac{3w^2-(u+v)^2}{3w^2-(u-v)^2}\right|\right),
\end{equation}
and
\begin{equation}
\mathcal{I}^A_{\mathrm{pc}1s}(u,v,y\rightarrow \infty)=0,
\end{equation}
thus
\begin{equation}\label{IPCs}
I^A_{\mathrm{PC}1s}(u,v,x \rightarrow \infty)=-\frac{3(u^2+v^2-3w^2)}{8u^3v^3}\left(4uv-(u^2+v^2-3w^2)\log\left|\frac{3w^2-(u+v)^2}{3w^2-(u-v)^2}\right|\right).
\end{equation}

As for $\mathcal{I}^A_{\mathrm{pc}1c}$, we have
\begin{equation}
\mathcal{I}^A_{\mathrm{pc}1c}(u,v,y\rightarrow 0)=0,
\end{equation}
and
\begin{equation}
\mathcal{I}^A_{\mathrm{pc}1c}(u,v,y\rightarrow \infty)=-\frac{3(u^2+v^2-3w^2)^2\pi}{8u^3v^3}\Theta(u+v-\sqrt{3}w),
\end{equation}
thus
\begin{equation}\label{IPCc}
I^A_{\mathrm{PC}1c}(u,v,x \rightarrow \infty)=-\frac{3(u^2+v^2-3w^2)^2\pi}{8u^3v^3}\Theta(u+v-\sqrt{3}w).
\end{equation}
From the above expressions \eqref{IPCs} and \eqref{IPCc}, if $w=1$, these two expressions are the same as those in GR \cite{Kohri:2018awv} except for an extra factor $1/2$ due to the definition of tensor perturbations.

\bibliographystyle{apsrev4-1}
\bibliography{main}

\end{document}